\documentclass[a4paper,12pt]{article}
\usepackage{graphicx,color,amsfonts,amstext,hyperref}

\newcommand{\TeXmacs}{{\TeX macs}}
\newcommand{\Macsyma}{\textit{\textsf{Macsyma}}}
\newcommand{\Maxima}{\textit{\textsf{Maxima}}}
\newcommand{\MuPAD}{\textit{\textsf{MuPAD}}}
\newcommand{\REDUCE}{\textit{\textsf{REDUCE}}}
\newcommand{\Math}{\textit{\textsf{Mathematica}}}
\newcommand{\Maple}{\textit{\textsf{Maple}}}

\newcommand{\red}{\color{red}}
\newcommand{\black}{\color{black}}
\newcommand{\blue}{\color{blue}}

\newcommand{\0}{{\char"5E}}
\newcommand{\Input}[2]{\texttt{#1\blue#2}}
\newcommand{\Output}[1]{\texttt{#1}}

\begin{document}

\title{\TeXmacs{} interfaces to \Maxima{}, \MuPAD{} and \REDUCE{}}
\author{A.~G.~Grozin\\
Budker Institute of Nuclear Physics, Novosibirsk 630090, Russia\\
A.G.Grozin@inp.nsk.su}
\date{}
\maketitle

\begin{abstract}
GNU \TeXmacs{} is a free wysiwyg word processor providing an excellent
typesetting quality of texts and formulae. It can also be used as an
interface to Computer Algebra Systems (CASs). In the present work, interfaces
to three general-purpose CASs have been implemented.
\end{abstract}

\section{\TeXmacs{}}\label{Sec:TeX}

GNU \TeXmacs{}~\cite{Ref:TeX} is a free (GPL) word processor which
\begin{itemize}
\item typesets texts and mathematical formulae with very high quality (like \LaTeX{}),
\item emphasizes the logical structure of a document rather than its appearance
(like \LaTeX{}),
\item is easy to use and intuitive (like typical wysiwyg word processors),
\item can be extended by a powerful programming language (like Emacs),
\item can include PostScript figures
(as well as other figures which can be converted to PostScript),
\item can export \LaTeX{}, and import \LaTeX{} and \texttt{html},
\item supports a number of languages based on Latin and Cyrillic alphabets.
\end{itemize}
It uses \TeX{} fonts both on screen and when printing documents.
Therefore, it is truly wysiwyg,
with equally good quality of on-screen and printed documents
(in contrast to LyX which uses X fonts on screen and calls \LaTeX{} for printing).
There is a similar commercial program called Scientific Workplace (for Windows).

\TeXmacs{} can also be used as an interface to any CAS which can generate \LaTeX{} output.
It renders \LaTeX{} formulae on the fly,
producing CAS output with highest typesetting quality
(better than, e.g., \Math{}, which uses fixed-width fonts for formula output).
A user can utilize editing facilities of \TeXmacs{}:
copy (a part of) a previous input into the new one, edit it and sent to the CAS,
copy a result derived using the CAS into a paper, etc.
In the present talk, I give some examples of using \Maxima{}, \MuPAD{} and \REDUCE{}
via \TeXmacs{}.
It is not my aim to describe these powerful and complex CASs;
I only show examples of typesetting produced by \TeXmacs{}.
This talk has been written in \TeXmacs{} and exported to \LaTeX{}.

\section{\Maxima{}}\label{Sec:Max}

\Macsyma{} is one of the oldest and most mature CASs.
It was developed at MIT during the end of sixties -- beginning of seventies.
Later, it was owned by various commercial companies.
Now it seems practically dead.

Fortunately, a free CAS \Maxima{} is now under GPL.
It is based upon the \Macsyma{} code base from seventies, with a number of later enhancements.
It incorporates a lot of mathematical knowledge, is stable and well tested.
From its very beginning, \Macsyma{} (and \Maxima{}) pays much attention
to mathematical correctness;
for example, if the form of an integral depends on the sign of a parameter,
it will ask the user about it, or use an assumption --
other systems only recently incorporated similar facilities.
It is an excellent platform for research projects, because it provides a solid foundation,
and it cannot vanish into thin air as commercial systems (e.g., \Macsyma{})
can do at any moment.
Its text-based interface now looks somewhat old-fashioned.
When combined with a nice graphical interface provided by \TeXmacs{},
it can compete with commercial CASs like \Math{} and \Maple{},
and even produce higher-quality output.
And it is difficult for commercial vendors to beat the price :--)

Here is a sample \Maxima{} session within \TeXmacs{}.

\begin{verbatim}
GCL (GNU Common Lisp) Version(2.4.0) Tue May 15 15:03:11 NOVST 2001
Licensed under GNU Library General Public License
Contains Enhancements by W. Schelter
Maxima 5.6 Tue May 15 15:03:08 NOVST 2001
(with enhancements by W. Schelter).
Licensed under the GNU Public License (see file COPYING)
\end{verbatim}

\noindent\Input{{\red (C1) {\black }}}{(x\02-y\02)/(x\02+y\02)+sin(alpha)\02;}

\noindent\Output{$\displaystyle{\text{\texttt{{\red (D1) {\black }}}} \frac{x^2 - y^2}{y^2 + x^2} +
\sin^2 \alpha}$}

\noindent\Input{{\red (C2) {\black }}}{expand((x+y-1)\05);}

\noindent\Output{$\displaystyle{\text{\texttt{{\red (D2) {\black }}}} y^5 + 5 x y^4 - 5 y^4 + 10 x^2
 y^3 - 20 x y^3 + 10 y^3 + 10 x^3 y^2 - 30 x^2 y^2 + 30 x y^2 - 10 y^2}$\\
$\displaystyle{{} + 5 x^4 y - 20 x^3 y + 30 x^2 y - 20 x y + 5 y + x^5 - 5 x^4 + 10 x^3 - 10 x^2 +
5 x - 1}$}

\newpage

\noindent\Input{{\red (C3) {\black }}}{solve(a*x\02+b*x+c,x);}

\noindent\Output{$\displaystyle{\text{\texttt{{\red (D3) {\black }}}} \left[x = - \frac{\sqrt{b^2 -
4 a c} + b}{2 a} , x = - \frac{b - \sqrt{b^2 - 4 a c}}{2 a} \right]}$}

\noindent\Input{{\red (C4) {\black }}}{integrate(sqrt(x\02+a),x);}

\noindent\Input{{\red ${\text{Is } a\text{ positive or negative?}}$ {\black
}}}{negative;}

\noindent\Output{$\displaystyle{\text{\texttt{{\red (D4) {\black }}}} \frac{a \log \left( 2
\sqrt{x^2 + a} + 2 x \right)}{2} + \frac{x \sqrt{x^2 + a}}{2}}$}

\noindent\Input{{\red (C5) {\black }}}{assume(a{>}0);}

\noindent\Output{$\displaystyle{\text{\texttt{{\red (D5) {\black }}}} \left[a > 0 \right]}$}

\noindent\Input{{\red (C6) {\black }}}{integrate(sqrt(x\02+a),x);}

\noindent\Output{$\displaystyle{\text{\texttt{{\red (D6) {\black }}}} \frac{a
\mathrm{\mathrm{ASINH}\;} \left( \frac{x}{\sqrt{a}} \right)}{2} + \frac{x
\sqrt{x^2 + a}}{2}}$}

\noindent\Input{{\red (C7) {\black }}}{integrate(exp(sin(x)),x,0,\%pi);}

\noindent\Output{$\displaystyle{\text{\texttt{{\red (D7) {\black }}}} \int_0^{\pi} e^{\sin
x}\; d x}$}

\noindent\Input{{\red (C8) {\black }}}{diff(f(x),x,2);}

\noindent\Output{$\displaystyle{\text{\texttt{{\red (D8) {\black }}}} \frac{d^2}{d x^2} f \left( x
\right)}$}

\noindent\Input{{\red (C9) {\black }}}{g:gamma(1+x);}

\noindent\Output{$\displaystyle{\text{\texttt{{\red (D9) {\black }}}} \Gamma \left( x + 1 \right)}$}

\noindent\Input{{\red (C10) {\black }}}{taylor(g,x,0,3);}

\noindent\Output{$\displaystyle{\text{\texttt{{\red (D10) {\black }}}} 1 - \gamma x + \frac{\left( 6
 \gamma^2 + \pi^2 \right) x^2}{12} - \frac{\left( 2 \gamma^3 + \pi^2
\gamma + 4 \zeta \left( 3 \right) \right) x^3}{12} + \cdots}$}

\noindent\Input{{\red (C11) {\black }}}{m:entermatrix(2,2);}

\noindent\Output{Is the matrix 1. Diagonal 2. Symmetric 3. Antisymmetric\\
4. General}

\noindent\Input{{\red Answer 1, 2, 3 or 4 : {\black }}}{4;}

\noindent\Input{{\red Row 1 Column 1: {\black }}}{a;}

\noindent\Input{{\red Row 1 Column 2: {\black }}}{b;}

\noindent\Input{{\red Row 2 Column 1: {\black }}}{c;}

\noindent\Input{{\red Row 2 Column 2: {\black }}}{d;}

\noindent\Output{Matrix entered.\\
$\displaystyle{\text{\texttt{{\red (D11) {\black }}}} \left( \begin{array}{cc} a & b\\ c &
d\\ \end{array} \right)}$}

\noindent\Input{{\red (C12) {\black }}}{m\0\0(-1);}

\noindent\Output{$\displaystyle{\text{\texttt{{\red (D12) {\black }}}} \left( \begin{array}{cc}
\frac{d}{a d - b c} & - \frac{b}{a d - b c} \\ - \frac{c}{a d - b c} &
\frac{a}{a d - b c} \\ \end{array} \right)}$}

\noindent\Input{{\red (C13) {\black }}}{fac(n):=if n=0 then 1 else n*fac(n-1);}

\noindent\Output{$\displaystyle{\text{\texttt{{\red (D13) {\black }}}}\mathrm{\mathrm{fac}} \left(
 n \right) : =\mathbf{if}\; n =
0\;\mathbf{then}\;
1\;\mathbf{else}\; n\,
\mathrm{fac} \left( n - 1 \right)}$}

\noindent\Input{{\red (C14) {\black }}}{trace(fac);}

\noindent\Output{$\displaystyle{\text{\texttt{{\red (D14) {\black }}}} \left[\mathrm{fac}
\right]}$}

\newpage

\noindent\Input{{\red (C15) {\black }}}{fac(5);}

\noindent\Output{$\text{1 Enter fac } \left[5 \right]$\\
$\text{\ 2 Enter fac } \left[4 \right]$\\
$\text{\ \ 3 Enter fac } \left[3 \right]$\\
$\text{\ \ \ 4 Enter fac } \left[2 \right]$\\
$\text{\ \ \ \ 5 Enter fac } \left[1 \right]$\\
$\text{\ \ \ \ \ 6 Enter fac } \left[0 \right]$\\
$\text{\ \ \ \ \ 6 Exit fac } 1$\\
$\text{\ \ \ \ 5 Exit fac } 1$\\
$\text{\ \ \ 4 Exit fac } 2$\\
$\text{\ \ 3 Exit fac } 6$\\
$\text{\ 2 Exit fac } 24$\\
$\text{1 Exit fac } 120$\\
${\text{\texttt{{\red (D15) {\black }}}} 120}$}

\noindent\Input{{\red (C16) {\black }}}{f(x):=block([a,z:0],a:x+1,a:a/z,a+1);}

\noindent\Output{$\displaystyle{\text{\texttt{{\red (D16) {\black }}}} f \left( x \right) :
=\mathbf{block}\; \left( \left[a , z : 0 \right] , a
: x + 1 , a : \frac{a}{z} , a + 1 \right)}$}

\noindent\Input{{\red (C17) {\black }}}{debugmode(true);}

\noindent\Output{$\displaystyle{\text{\texttt{{\red (D17) {\black }}}}\mathbf{true}}$}

\noindent\Input{{\red (C18) {\black }}}{f(u);}

\noindent\Output{Division by 0\\
-- an error. Entering the Maxima Debugger dbm f(x=u)}

\noindent\Input{{\red (dbm:1) {\black }}}{a;}

\noindent\Output{$\displaystyle{u + 1}$}

\noindent\Input{{\red (dbm:1) {\black }}}{z;}

\noindent\Output{$\displaystyle{0}$}

\noindent\Input{{\red (dbm:1) {\black }}}{:q}

\noindent\Input{{\red (C19) {\black }}}{plot2d(sin(x)/x,[x,-10,10]);}

\noindent\Output{$\displaystyle{\text{\texttt{{\red (D19) {\black }}}} 0}$}

\noindent\Input{{\red (C20) {\black }}}{f(x,y):=sin(sqrt(x\02+y\02))/sqrt(x\02+y\02);}

\noindent\Output{$\displaystyle{\text{\texttt{{\red (D20) {\black }}}} f \left( x , y \right) : =
\frac{\sin \sqrt{x^2 + y^2}}{\sqrt{x^2 + y^2}}}$}

\noindent\Input{{\red (C21) {\black }}}{plot3d(f(x,y),[x,-10,10],[y,-10,10]);}

\noindent\Output{$\displaystyle{\text{\texttt{{\red (D21) {\black }}}} 0}$}

\begin{figure}[ht]
\begin{flushleft}
\includegraphics[width=7cm]{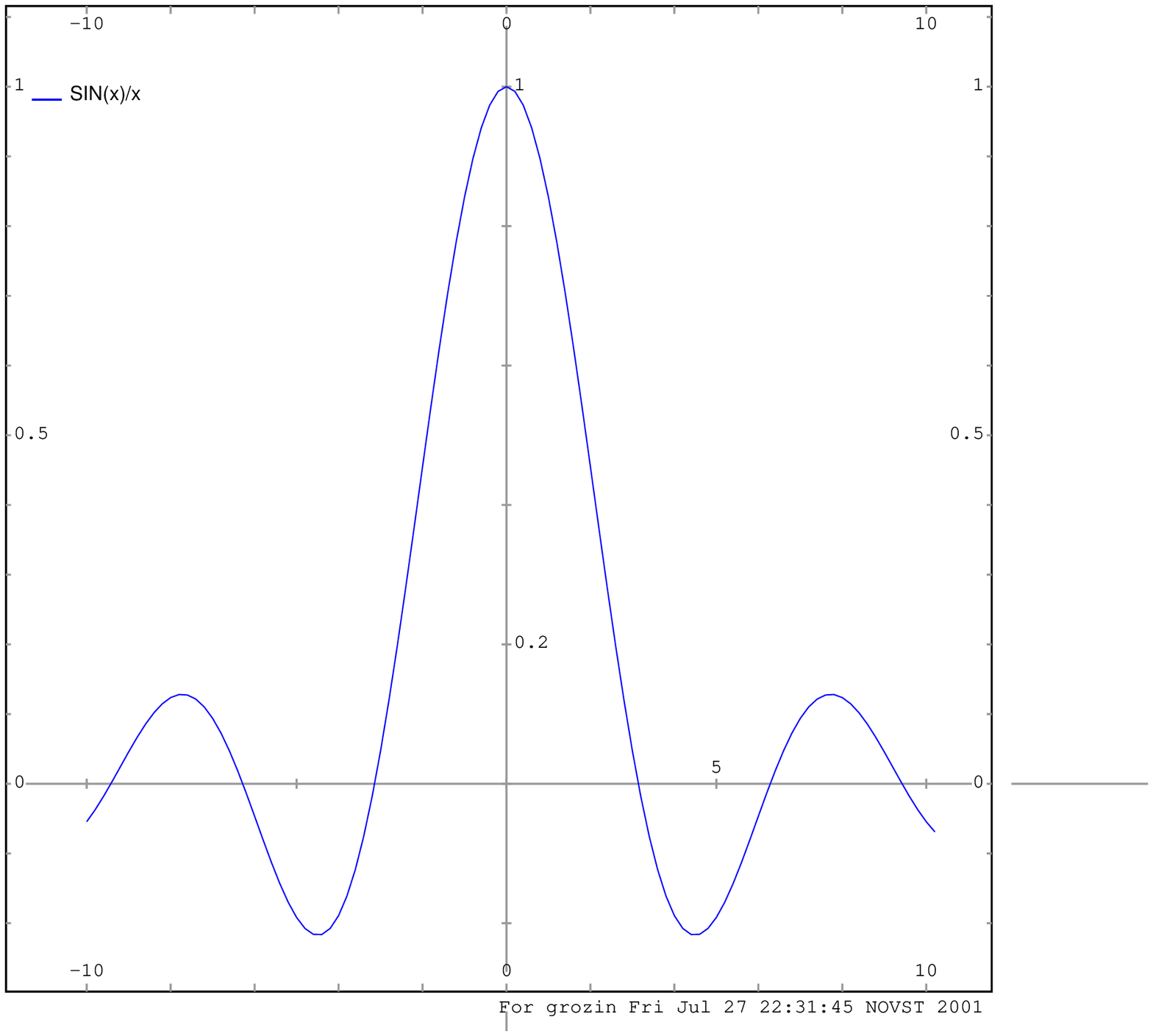}
\hfill
\includegraphics[width=7cm]{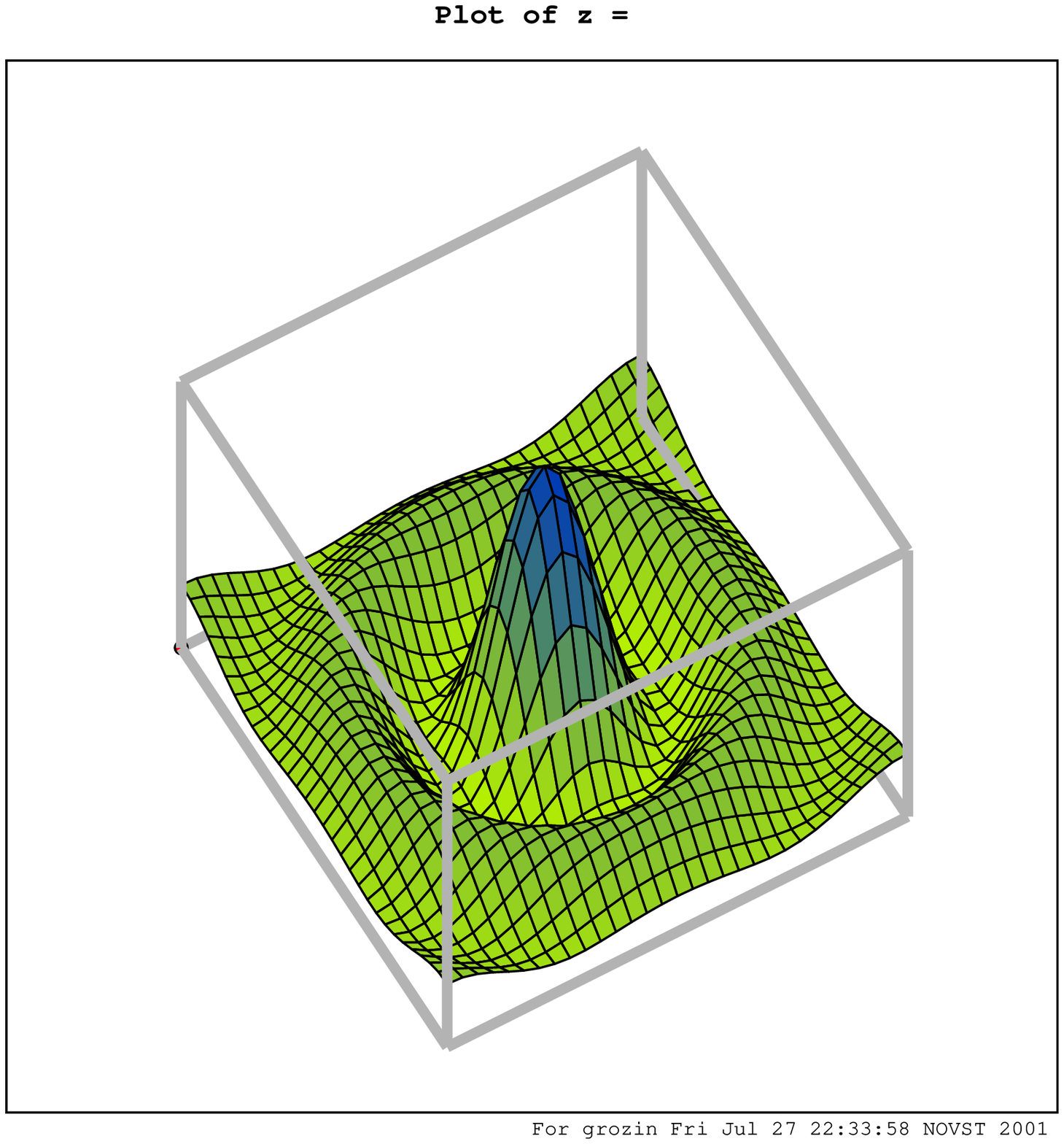}
\end{flushleft}
\caption{\Maxima{} plots}\label{Fig:Max}
\end{figure}

Plots appear in separate windows (Fig.~\ref{Fig:Max}).
When the mouse is over such a window,
its coordinates are continuously displayed at the upper left corner
(in the 3d case, also $z$ of the surface at the mouse position $( x , y )$ is shown).
When the mouse is at the upper left corner, a menu appears.
It allows the user to control the plot: zoom, rotate (in the 3d case), print,
save as PostScript, etc.

The toolbar icon showing the question mark shows the \Maxima{} documentation.
The \Maxima{} manual is in html; it is imported into \TeXmacs{}
and shown in a new buffer.
Hyperlinks work with double click.
It is easy to arrange things in such a way
that you do calculations in one \TeXmacs{} window,
and read the manual in another window.

\section{\MuPAD{}}\label{Sec:Mu}

\MuPAD{}~\cite{Ref:MuPAD} is the most recent addition to the family of universal CASs.
It is being developed at University of Paderborn,
and commercially distributed by SciFace.
In some cases, it can be obtained free of charge (see their web site),
but it is not free software.
It is rather similar to \Maple{}, but designed from scratch,
and some new fundamental ideas were incorporated.
The library is not so extensive as those of older systems,
but \MuPAD{} is progressing fast.
Its interface is text-based (except the Windows version).
Therefore, adding a high-quality graphical formula output provided by \TeXmacs{} is useful.

Here is a sample \MuPAD{} session within \TeXmacs{}.

\begin{verbatim}
   *----*    MuPAD 2.0.0 -- The Open Computer Algebra System
  /|   /|
 *----* |    Copyright (c)  1997 - 2000  by SciFace Software
 | *--|-*                   All rights reserved.
 |/   |/
 *----*      Licensed to:   Andrey Grozin
\end{verbatim}

\noindent\Input{{\red $\gg $ {\black }}}{(x\02-y\02)/(x\02+y\02)+sin(alpha)\02}

\noindent\Output{$\displaystyle{\sin \left( \alpha \right)^2 + \frac{\left( x^2 - y^2
\right)}{\left( x^2 + y^2 \right)}}$}

\newpage

\noindent\Input{{\red $\gg $ {\black }}}{expand((x+y-1)\05)}

\noindent\Output{$\displaystyle{5\, x + 5\, y - 20\,
x\, y - 10\, x^2 + 10\, x^3 -
10\, y^2 - 5\, x^4 + 10\, y^3 + x^5
- 5\, y^4 + y^5 + 30\, x\, y^2}$\\
$\displaystyle{{} + 30\, x^2\, y - 20\,
x\, y^3 - 20\, x^3\, y +
5\, x\, y^4 + 5\,
x^4\, y - 30\, x^2\, y^2 +
10\, x^2\, y^3 + 10\,
x^3\, y^2 - 1}$}

\noindent\Input{{\red $\gg $ {\black }}}{solve(a*x\02+b*x+c=0,x)}

\noindent\Output{$\displaystyle{\left\{
\begin{array}{cc}
\mathbb{C} & \text{if}\, a = 0 \wedge b = 0 \wedge c = 0\\
\left\{\right\} & \text{if}\, a = 0 \wedge b = 0 \wedge c \neq 0\\
\left\{- \frac{c}{b} \right\} & \text{if}\, a = 0 \wedge b \neq 0\\
\left\{\frac{- \frac{b}{2} -
\frac{\sqrt{b^2 - 4\, a\, c}}{2}}{a} ,
\frac{\frac{\sqrt{b^2 - 4\, a\, c}}{2} -
\frac{b}{2}}{a} \right\} & \text{if}\, a \neq 0
\end{array}
\right.}$}

\noindent\Input{{\red $\gg $ {\black }}}{int(sqrt(x\02+a),x)}

\noindent\Output{$\displaystyle{\frac{x\, \sqrt{a + x^2}}{2} +
\frac{a\,\text{ln} \left( x + \sqrt{a + x^2} \right)}{2}}$}

\noindent\Input{{\red $\gg $ {\black }}}{i1:=int(exp(sin(x)),x=0..PI); float(i1)}

\noindent\Output{$\displaystyle{\int_0^{\pi}\text{exp} \left( \sin \left( x \right) \right) d x}$\\
${6.208758036}$}

\noindent\Input{{\red $\gg $ {\black }}}{diff(f(x),x,x)}

\noindent\Output{$\displaystyle{\frac{\partial^2}{\partial x^2} f \left( x \right)}$}

\noindent\Input{{\red $\gg $ {\black }}}{g:=gamma(1+x)}

\noindent\Output{$\displaystyle{\gamma \left( x + 1 \right)}$}

\noindent\Input{{\red $\gg $ {\black }}}{series(g,x=0,4)}

\noindent\Output{$\displaystyle{1 - x\, \gamma + x^2\, \left( \frac{\pi^2}{12}
+ \frac{\gamma^2}{2} \right) + x^3\, \left( -
\frac{\zeta \left( 3 \right)}{3} - \frac{\gamma^3}{6} - \frac{\pi^2\, \gamma}{12}
\right) + O \left( x^4 \right)}$}

\noindent\Input{{\red $\gg $ {\black }}}{M:=matrix([[a,b],[c,d]]); 1/M}

\noindent\Output{$\displaystyle{\left( \begin{array}{cc} a & b\\ c & d\\ \end{array} \right)}$\\
$\displaystyle{\left( \begin{array}{cc} - \frac{d}{b\, c - a\,
d} & \frac{b}{b\, c - a\, d} \\
\frac{c}{b\, c - a\, d} & -
\frac{a}{b\, c - a\, d} \\ \end{array} \right)}$}

\noindent\Input{{\red $\gg $ {\black }}}{plotfunc2d(sin(x)/x,x=-10..10)}

\noindent\Output{Warning: Dumb terminal: Plot data saved in binary file save.mp\\
{}[plot]; during evaluation of 'plot2d'}

\noindent\Input{{\red $\gg $ {\black }}}{f:=(x,y)-{>}sin(sqrt(x\02+y\02))/sqrt(x\02+y\02)}

\noindent\Output{${\text{(x, y) -{>}sin(sqrt(x\02+ y\02))/sqrt(x\02+ y\02)}}$}

\noindent\Input{{\red $\gg $ {\black }}}{plotfunc3d(f(x,y),x=-10..10,y=-10..10)}

\noindent\Output{Warning: Dumb terminal: Plot data saved in binary file save.mp\\
{}[plot]; during evaluation of 'plot3d'}

\noindent\Input{{\red $\gg $ {\black }}}{quit}

\noindent\Output{{\red The end{\black }}}

\begin{figure}[ht]
\begin{flushleft}
\includegraphics[width=7cm]{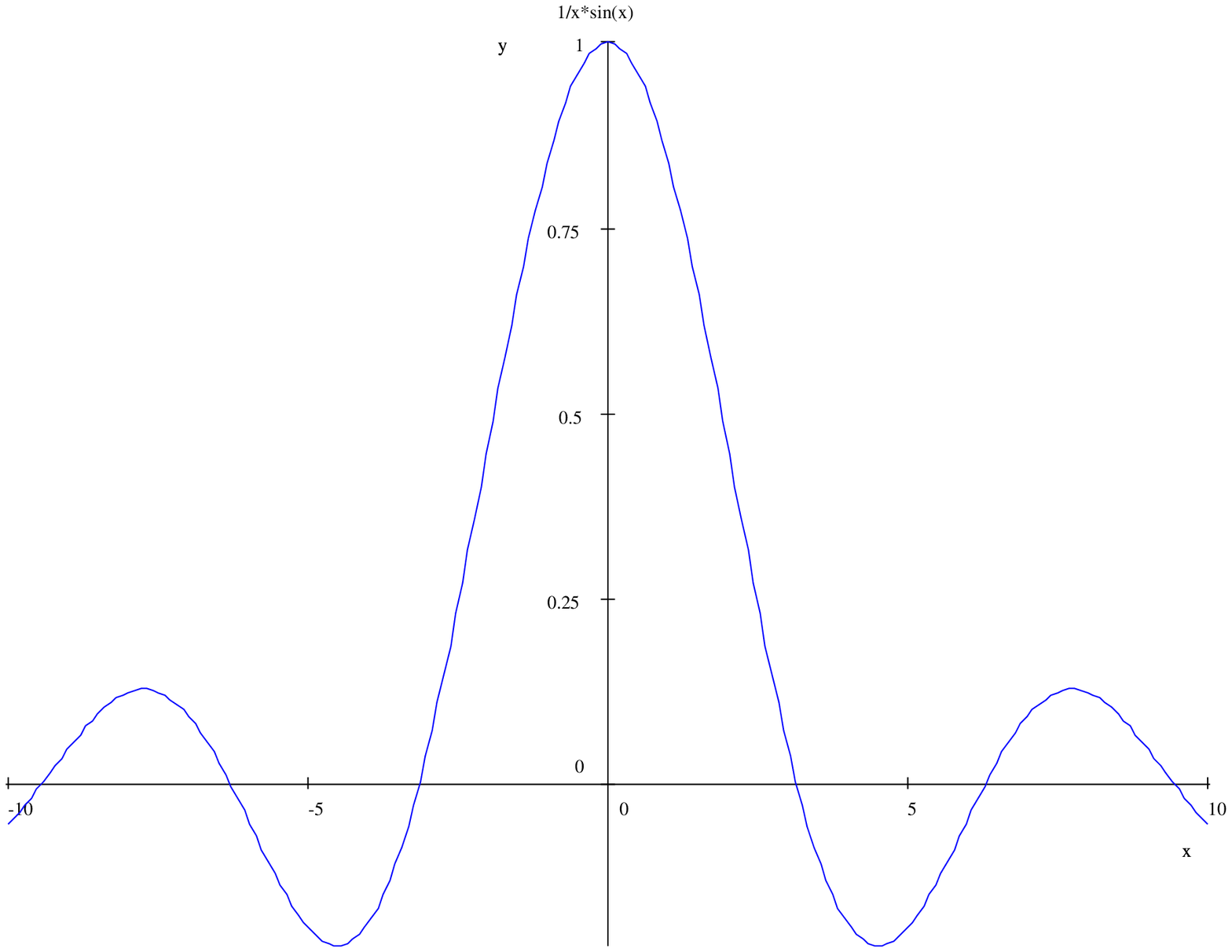}
\hfill
\includegraphics[width=7cm]{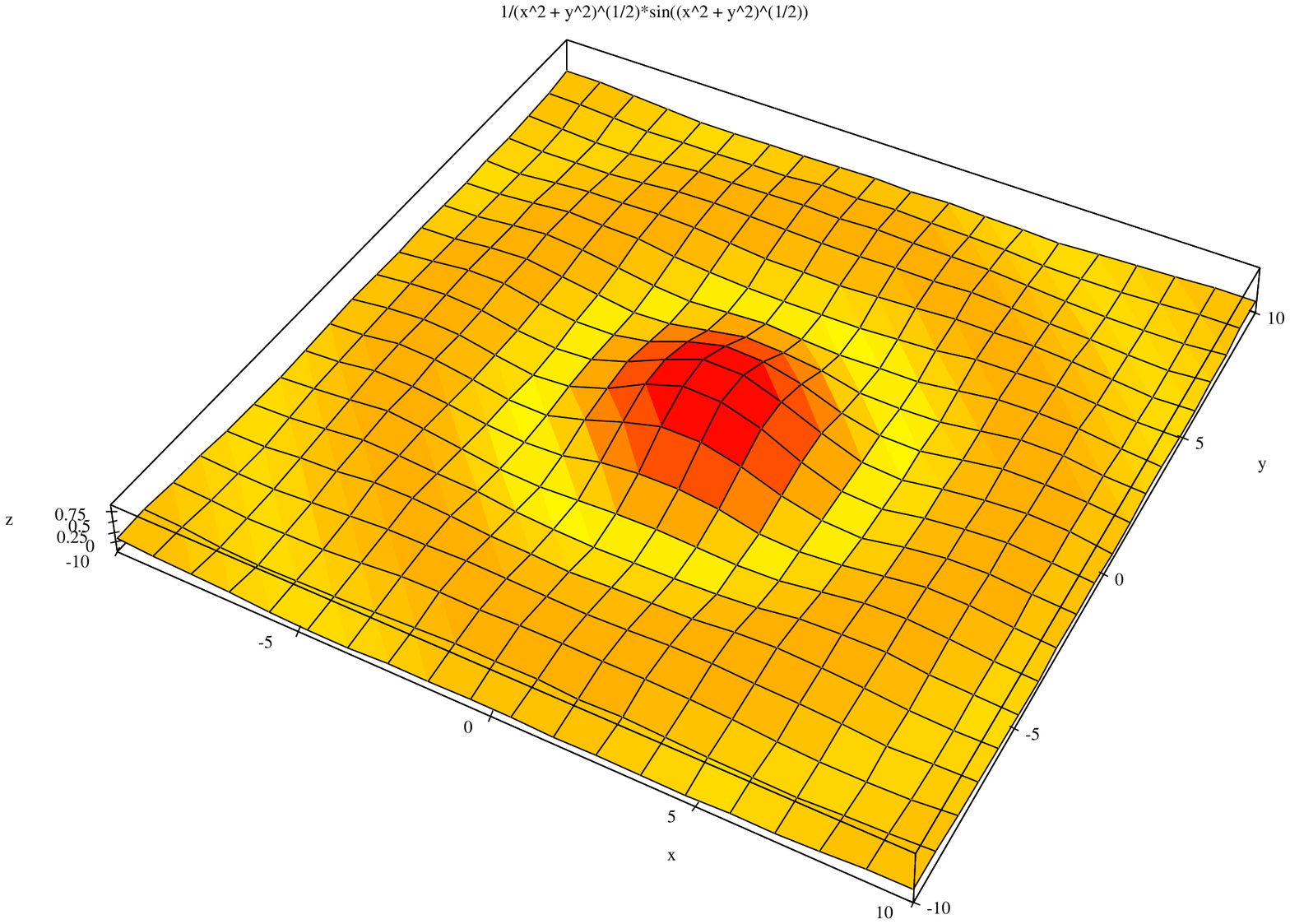}
\end{flushleft}
\caption{\MuPAD{} plots}\label{Fig:MuPAD}
\end{figure}

Plots appear in separate windows (Fig.~\ref{Fig:MuPAD}).
Spurious warnings about dumb terminal may be ignored.
Plots are displayed by the program \texttt{vcam} which is distributed with \MuPAD{}.
They can be controlled (and saved to PostScript files) via menus.

The question mark icon opens the help menu.
\MuPAD{} documentation is extensive and high-quality.
It is written in an extension of \LaTeX{} with hyperlinks,
and displayed by an extended \texttt{dvi} viewer distributed with \MuPAD{}.
Therefore, it has high typesetting quality.

\section{\REDUCE{}}\label{Sec:Red}

\REDUCE{} is one of the older CASs (it was somewhat influenced by \Macsyma{}).
It is a commercial system.
It is stable and efficient,
and can solve larger problems in a given memory than, say, \Math{}.
Its text-based interface looks old-fashioned;
it has also an X interface,
which provides 2-dimensional formula output (not of a very high quality)
and a convenient on-line help.

Here is a sample \REDUCE{} session within \TeXmacs{}.

\begin{verbatim}
Loading image file :/opt/reduce/lisp/psl/linux/red/reduce.img 
REDUCE 3.7, 15-Apr-1999, patched to 14-Jun-2001 ...
\end{verbatim}

\noindent\Input{{\red 1: {\black }}}{(x\02-y\02)/(x\02+y\02)+sin(alpha)\02;}

\noindent\Output{$\displaystyle{\frac{\sin \left( \alpha \right)^2\,
x^2\, +\, \sin \left( \alpha \right)^2\, y^2\, +\, x^2\,
-\, y^2}{x^2\, +\, y^2}}$}

\noindent\Input{{\red 2: {\black }}}{(x+y-1)\05;}

\noindent\Output{$\displaystyle{x^5\, +\, 5\,
x^4\, y\, -\, 5\,
x^4\, +\, 10\, x^3\,
y^2\, -\, 20\, x^3\,
y\, +\, 10\, x^3\,
+\, 10\, x^2\, y^3\,
-\, 30\, x^2\, y^2\,
+\, 30\, x^2\, y\,}$\\
$\displaystyle{{}-\, 10\, x^2\, +\,
5\, x\, y^4\, -\,
20\, x\, y^3\, +\,
30\, x\, y^2\, -\,
20\, x\, y\, +\,
5\, x\, +\, y^5\,
-\, 5\, y^4\, +\,
10\, y^3\,}$\\
$\displaystyle{{}-\, 10\,y^2\, +\, 5\, y\,-\, 1}$}

\noindent\Input{{\red 3: {\black }}}{solve(a*x\02+b*x+c=0,x);}

\noindent\Output{$\displaystyle{\left\{x = \frac{\sqrt{\, -\,
4\, a\, c\, +\, b^2}
\, -\, b}{2\, a} \,
,\,\, x = \frac{\,
-\, \left( \sqrt{\, -\,
4\, a\, c\, +\, b^2}
\, +\, b \right)}{2\, a} \right\}}$}

\noindent\Input{{\red 4: {\black }}}{int(sqrt(x\02+a),x);}

\noindent\Output{$\displaystyle{\frac{\sqrt{a\, +\, x^2}
\, x\, +\, \log \left(
\frac{\sqrt{a\, +\, x^2} \,
+\, x}{\sqrt{a}} \right) \, a}{2}}$}

\noindent\Input{{\red 5: {\black }}}{int(exp(sin(x)),x);}

\noindent\Output{$\displaystyle{\int e^{\sin \left( x \right)}\, d\,
x}$}

\noindent\Input{{\red 6: {\black }}}{df(f(x),x,2);}

\noindent\Input{{\red Declare f operator ? {\black }}}{y}

\noindent\Output{$\displaystyle{\frac{\partial^2\, f \left( x \right)}{\partial
\, x^2}}$}

\noindent\Input{{\red 7: {\black }}}{taylor(sin(x),x,0,10);}

\noindent\Output{$\displaystyle{x\, -\, \frac{1}{6} \,
x^3\, +\, \frac{1}{120} \,
x^5\, -\, \frac{1}{5040} \,
x^7\, +\, \frac{1}{362880} \,
x^9\, +\, O \left( x^{11} \right)}$}

\noindent\Input{{\red 8: {\black }}}{m:=mat((a,b),(c,d));}

\noindent\Output{$\displaystyle{m : = \left( \begin{array}{cc} a & b\\ c & d\\ \end{array}
\right)}$}

\noindent\Input{{\red 9: {\black }}}{1/m;}

\noindent\Output{$\displaystyle{\left( \begin{array}{cc} \frac{d}{a\,
d\, -\, b\, c} &
\frac{\, -\, b}{a\,
d\, -\, b\, c} \\
\frac{\, -\, c}{a\,
d\, -\, b\, c} &
\frac{a}{a\, d\, -\,
b\, c} \\ \end{array} \right)}$}

\noindent\Input{{\red 10: {\black }}}{plot(sin(x)/x,x=(-10 .. 10));}

\noindent\Input{{\red 11: {\black }}}{procedure f(x,y);
sin(sqrt(x\02+y\02))/sqrt(x\02+y\02);}

\noindent\Output{${f}$}

\noindent\Input{{\red 12: {\black }}}{plot(f(x,y),x=(-10 .. 10),y=(-10 ..
10),hidden3d,points=40);}

\noindent\Input{{\red 13: {\black }}}{bye;}

\noindent\Output{${Q u i t t i n g}$\\
{\red The end{\black }}}

\begin{figure}[ht]
\begin{flushleft}
\includegraphics[width=7cm]{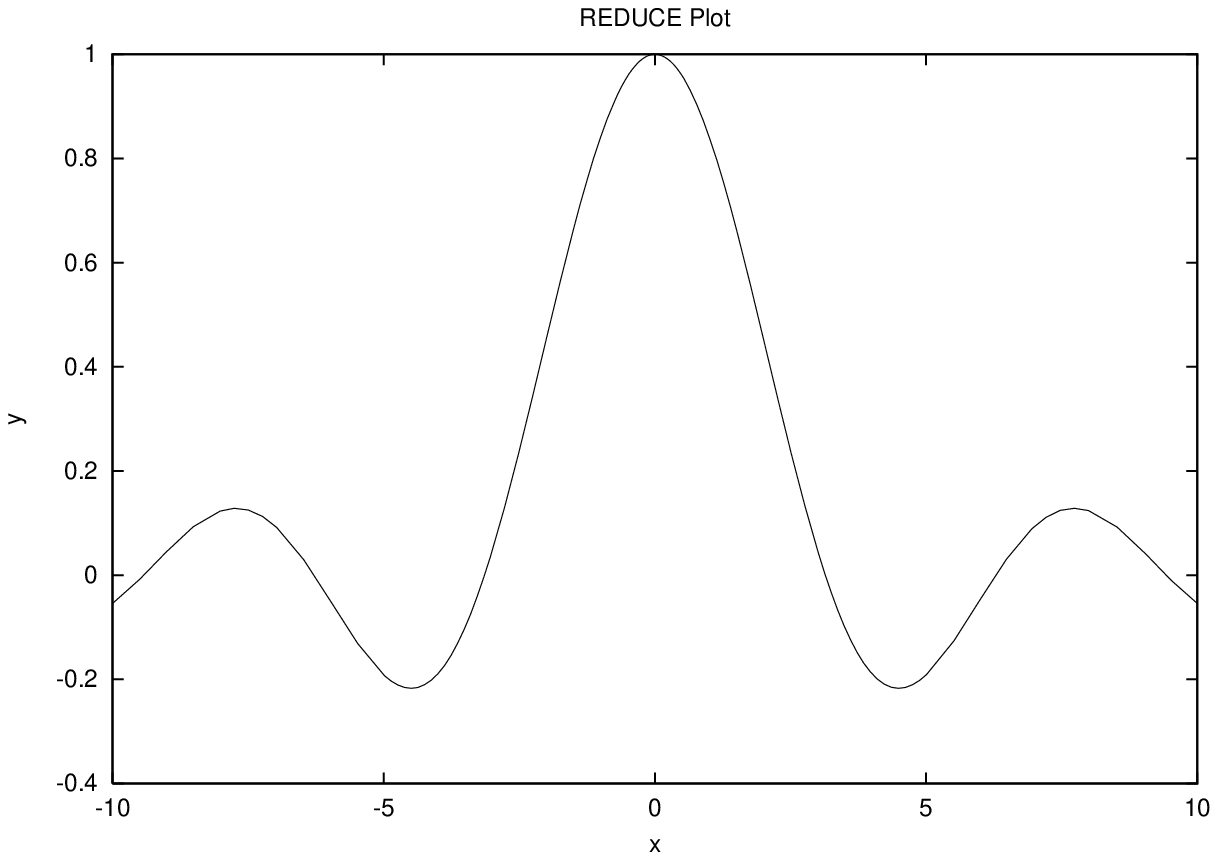}
\hfill
\includegraphics[width=7cm]{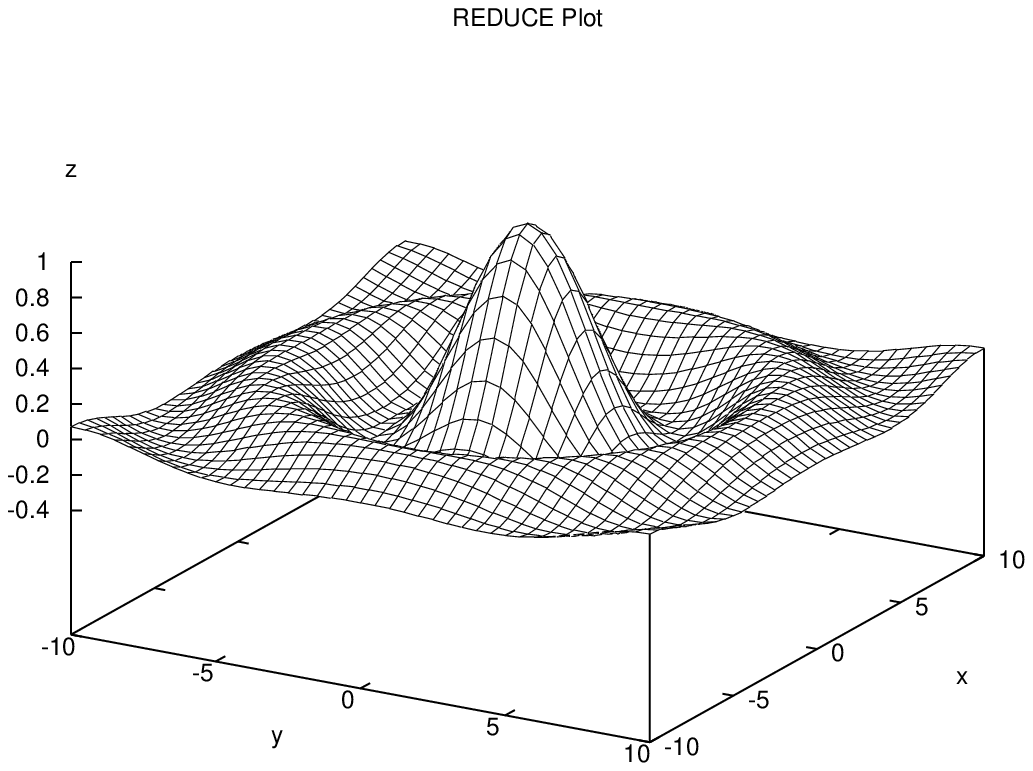}
\end{flushleft}
\caption{\REDUCE{} plots}\label{Fig:Red}
\end{figure}

Plots appear in separate windows, they are displayed by \texttt{gnuplot}
(Fig.~\ref{Fig:Red}).
Unfortunately, it is not possible to control them interactively,
one has to use options in the \texttt{plot} procedure.
In order to save a plot to a PostScript file, the options
\begin{quote}
\begin{verbatim}
terminal="postscript eps",output="filename.eps"
\end{verbatim}
\end{quote}
are used.

The question mark icon displays help menu.
\REDUCE{} manual (written in \LaTeX{}) is imported into \TeXmacs{}.
Some \LaTeX{} constructs are not handled correctly, but, nevertheless,
the manual is quite readable.

All interfaces described in this talk are preliminary, and require more work.
It is not difficult to implement \TeXmacs{} interfaces with more CASs.
\TeXmacs{} progresses fast; in the future, it can become a complete scientist's work
place, suitable both for writing articles and for doing calculations using
various external systems, within the same comfortable environment.

I am grateful
to Joris van der Hoeven for numerous discussions about \TeXmacs{} and CAS interfaces;
to William Schelter for his great help with \Maxima{} and its \LaTeX{} generation;
to Ralf Hillebrand for similar help with \MuPAD{}
and for providing an improved \LaTeX{} generation library;
to Winfried Neun for useful discussions about \REDUCE{}.

\end{document}